\documentclass[acmsmall]{acmart}

\AtBeginDocument{%
  }

\newcommand{\tool}{\texttt{TraceDroid}}
\usepackage[linesnumbered,ruled,vlined]{algorithm2e}

\usepackage{float}
\usepackage{threeparttable}
\usepackage{booktabs}
\usepackage{multirow}
\usepackage{epstopdf}
\usepackage{hyperref}
\usepackage{listings}
\usepackage{fancybox}
\usepackage{graphicx}
\usepackage{subcaption}
\usepackage{paralist}
\usepackage{ragged2e}
\usepackage{enumitem}

\usepackage{marvosym}
\usepackage{multirow}
\usepackage{epstopdf}
\usepackage{hyperref}
\usepackage{listings}
\usepackage{fancybox}
\usepackage{graphicx}
\usepackage{subcaption}
\usepackage{subfloat}
\usepackage{cleveref}
\usepackage{paralist}
\usepackage{ragged2e}
\usepackage{enumitem}
\usepackage{pythonhighlight}
\usepackage{algpseudocode}

\usepackage[normalem]{ulem} 
\newcommand\hlg{\bgroup\markoverwith
  {\textcolor{green!10}{\rule[-.5ex]{2pt}{2.5ex}}}\ULon}
\newcommand\hlb{\bgroup\markoverwith
  {\textcolor{blue!10}{\rule[-.5ex]{2pt}{2.5ex}}}\ULon}
\newcommand\hlr{\bgroup\markoverwith
  {\textcolor{red!10}{\rule[-.5ex]{2pt}{2.5ex}}}\ULon}

\usepackage{listings}

\newif\ifFirstMintedPart


\usepackage[table]{xcolor}
\usepackage{soul}
\usepackage{float}
\usepackage{threeparttable}
\usepackage{booktabs}
\usepackage{multirow}
\usepackage{epstopdf}
\usepackage{hyperref}
\usepackage{listings}
\usepackage{fancybox}
\usepackage{graphicx}
\usepackage{subcaption}
\usepackage{paralist}
\usepackage{ragged2e}
\usepackage{enumitem}
\usepackage{balance}

\usepackage{marvosym}
\usepackage{multirow}
\usepackage{epstopdf}
\usepackage{hyperref}
\usepackage{listings}
\usepackage{fancybox}
\usepackage{graphicx}
\usepackage{subcaption}
\usepackage{subfloat}
\usepackage{cleveref}
\usepackage{paralist}
\usepackage{ragged2e}
\usepackage{enumitem}
\usepackage{pythonhighlight}
\usepackage{algpseudocode}

\usepackage{minted}
\usepackage{caption}
\usepackage{xparse}

\newminted{c}{
    linenos, 
    frame=lines, 
    tabsize=4, 
    autogobble,
}

\newminted{java}{
    linenos, 
    frame=lines, 
    tabsize=4, 
    autogobble,
}

\usepackage{listings}

\usepackage[framemethod=TikZ]{mdframed}
\usepackage{tcolorbox}
\makeatletter
\newcommand{\mybox}[1]{%
  \setbox0=\hbox{#1}%
  \setlength{\@tempdima}{\dimexpr\wd0+13pt}%
  \begin{tcolorbox}[boxrule=0.5pt, colback=white, arc=4pt,
      left=6pt,right=6pt,top=6pt,bottom=6pt,boxsep=0pt]
    #1
  \end{tcolorbox}
}
\usepackage{tikz}
\usepackage{pgfplots}
\usetikzlibrary{pgfplots.statistics,calc}
\usepackage{array}
\usepackage{centernot}
\usepackage{xspace}
\usepackage{url}
\usepackage{verbatim}
\usepackage{wrapfig}
\usepackage{tabularx}
\usepackage{bbding}
\usepackage{pifont}
\usepackage{wasysym}
\usepackage{amssymb}

\setcopyright{cc}
\setcctype{by}
\acmDOI{10.1145/3808158}
\acmYear{2026}
\acmJournal{PACMSE}
\acmVolume{3}
\acmNumber{FSE}
\acmArticle{FSE151}
\acmMonth{7}
\acmSubmissionID{fse26mainb-p844-p}
\received{2025-09-12}
\received[accepted]{2026-03-24}

\begin{document}

\title{From Suspicious Signals to Crashes: Guiding Bug-Driven GUI Testing via Code-Inspired Tracing}

\author{Mengzhuo Chen}
\authornote{Both authors contributed equally to this research.}
\orcid{0009-0006-4397-750X}
\affiliation{%
  \department{Institute of Software and State Key Laboratory of Intelligent Game}
  \institution{Chinese Academy of Sciences}
  \country{China}
}
\affiliation{%
  \institution{University of Chinese Academy of Sciences}
  \city{Beijing}
  \country{China}
}
\email{mengzhuo.happy@gmail.com}

\author{Zhe Liu}
\authornotemark[1]
\orcid{0000-0001-9709-8275}
\affiliation{%
  \department{Institute of Software and State Key Laboratory of Intelligent Game}
  \institution{Chinese Academy of Sciences}
  \country{China}
}
\affiliation{%
  \institution{University of Chinese Academy of Sciences}
  \city{Beijing}
  \country{China}
}
\email{liuzhe181@mails.ucas.edu.cn}

\author{Chunyang Chen}
\orcid{0000-0003-2011-9618}
\affiliation{%
  \institution{Technical University of Munich}
  \city{Munich}
  \country{Germany}
}
\email{chun-yang.chen@tum.de}

\author{Junjie Wang}
\authornote{Corresponding author.}
\orcid{0000-0002-9941-6713}
\affiliation{%
  \department{Institute of Software and State Key Laboratory of Intelligent Game}
  \institution{Chinese Academy of Sciences}
  \country{China}
}
\affiliation{%
  \institution{University of Chinese Academy of Sciences}
  \city{Beijing}
  \country{China}
}
\email{junjie@iscas.ac.cn}

\author{Yuekai Huang}
\orcid{0000-0003-2762-9223}
\affiliation{%
  \department{Institute of Software and State Key Laboratory of Intelligent Game}
  \institution{Chinese Academy of Sciences}
  \country{China}
}
\affiliation{%
  \institution{University of Chinese Academy of Sciences}
  \city{Beijing}
  \country{China}
}
\email{boyu_wu2021@163.com}

\author{Boyu Wu}
\orcid{0009-0001-9285-3419}
\affiliation{%
  \department{Institute of Software and State Key Laboratory of Intelligent Game}
  \institution{Chinese Academy of Sciences}
  \country{China}
}
\affiliation{%
  \institution{University of Chinese Academy of Sciences}
  \city{Beijing}
  \country{China}
}
\email{huangyuekai18@mails.ucas.ac.cn}

\author{Jun Hu}
\orcid{0009-0005-1530-7499}
\affiliation{%
  \department{Institute of Software and State Key Laboratory of Intelligent Game}
  \institution{Chinese Academy of Sciences}
  \country{China}
}
\affiliation{%
  \institution{University of Chinese Academy of Sciences}
  \city{Beijing}
  \country{China}
}
\email{hujun@iscas.ac.cn}

\author{Qing Wang}
\authornotemark[2]
\orcid{0000-0002-2618-5694}
\affiliation{%
  \department{Institute of Software and State Key Laboratory of Intelligent Game}
  \institution{Chinese Academy of Sciences}
  \country{China}
}
\affiliation{%
  \institution{University of Chinese Academy of Sciences}
  \city{Beijing}
  \country{China}
}
\email{wq@iscas.ac.cn}









\begin{abstract}
Mobile applications frequently suffer from crash bugs that are triggered under specific GUI interaction sequences. Existing automated GUI testing techniques mainly emphasize increasing coverage through diverse exploration strategies, but they often fail to reach the precise interaction contexts that lead to crashes, resulting in low bug detection efficiency. 
This paper proposes {\tool}, a novel automated GUI testing approach that leverages suspicious code-level signals to guide dynamic exploration. Instead of treating static analysis as an independent detection method, {\tool} uses heuristic rules distilled from real crash reports to detect suspicious code segments, associate them with GUI widgets, and collect code-level interaction signals. It then constructs the Activity Transition Graph (ATG), performs rough path generation, and employs LLM-based executable path completion to produce a set of suspicious paths. Finally, {\tool} executes these paths through global path planning, local path generation, and execution-aware monitoring to efficiently expose crashes.  
We evaluate {\tool} on 70 real crash bugs across 42 open-source apps, comparing it with 15 state-of-the-art baselines. {\tool} achieves the best performance, with a recall of 77\%, exceeding the best baseline by 28\%, while maintaining comparable or higher coverage. Furthermore, {\tool} successfully detects 21 previously unknown crash bugs in 116 popular Google Play apps, of which 15 have been fixed and 6 confirmed by developers, demonstrating its effectiveness in real-world scenarios.
\end{abstract}

\ccsdesc[500]{Software and its engineering~Software testing and debugging}





\keywords{Mobile App, Large language model, GUI testing}

\maketitle

\section{Introduction}
\label{sec:Introduction}
Mobile applications have become an integral part of daily life, and ensuring their reliability has become a critical concern for both developers and researchers. Software crash bugs often have a huge impact on our software usage, which can abruptly terminate an app's execution, potentially leading to data loss, broken workflows, or negative user experiences. Research shows that crash bugs often occur under specific interaction conditions, many of which require a precise sequence of GUI operations to activate, making crash bugs inherently difficult to uncover. To detect crash bugs, researchers and practitioners typically employ automated GUI testing techniques, which simulate user interactions and observe application behavior during execution to capture crash bugs.

\begin{figure*}[htb]
\centering
\includegraphics[width=13.7cm]{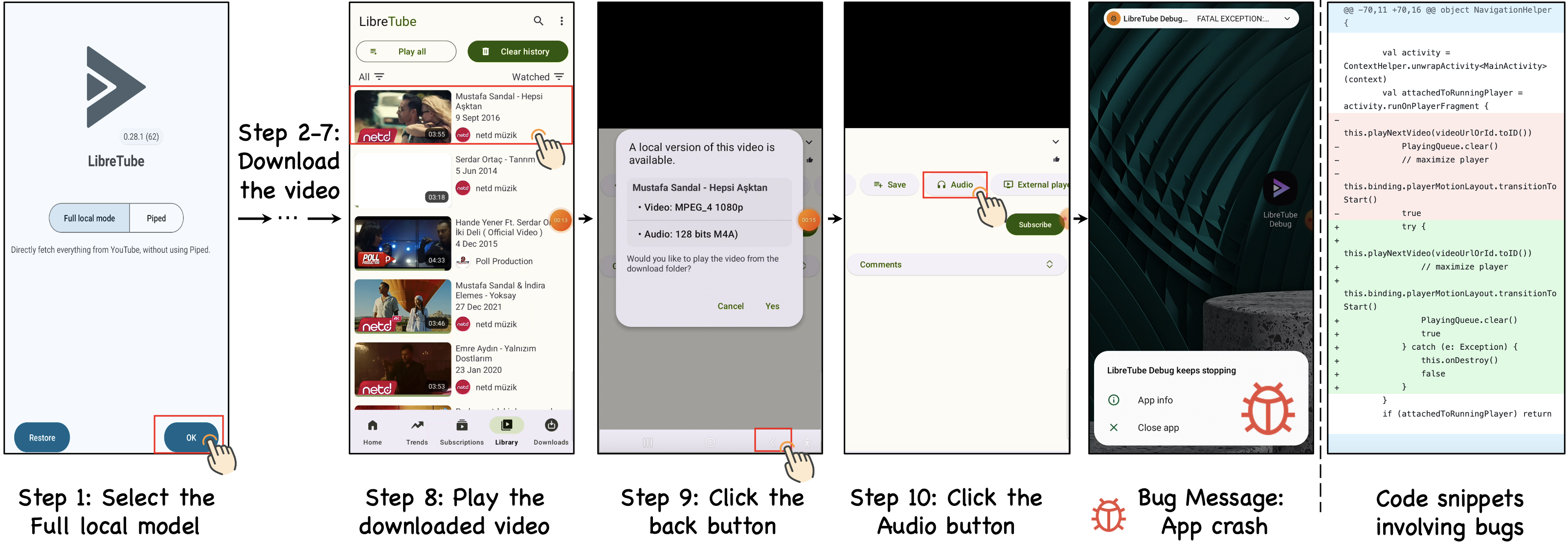}
\vspace{-0.1in}
\caption{Example of a crash bug and its bug code}
\label{fig:example}
\vspace{-0.2in}
\end{figure*}

In recent years, automated GUI testing has continuously evolved with more sophisticated exploration strategies. Model-based methods~\cite{mirzaei2016reducing,yang2018static,yang2013grey,zeng2016automated,mao2016sapienz,su2017guided,dong2020time,gu2019practical,wang2020combodroid,liu2023ex} aiming to increase activity or code coverage under the assumption that more coverage correlates with more bugs found. Learning-based~\cite{li2019humanoid,pan2020reinforcement} and LLM-driven methods~\cite{liu2024make,yang2023appagent,yoon2023autonomous,hu2024auitestagent,taeb2024axnav,lu2025uxagent,li2024test} imitate human interaction patterns to generate more diverse and realistic action sequences. These methods expand coverage and interaction diversity, but they are not explicitly designed to prioritize crash-prone behaviors. As a result, crash bugs that require specific or long interaction sequences are often undetected. As shown in Fig.\ref{fig:example}, in the open-source app \textit{LibreTube}\cite{Libretube}, a crash is triggered after following a relatively long path of at least ten steps, involving multiple functionalities such as downloading a video, switching between playback modes, navigating back, and finally clicking the audio conversion button. The crash does not occur if the user selects the ``Yes'' option in the intermediate dialog. Instead, it arises when the user dismisses the dialog via the back button and then proceeds to the audio function. Although each step is valid on its own, the large number of possible operations makes it infeasible to cover all combinations. This also requires GUI testing not only to achieve coverage, but also to determine which paths should be prioritized for exploration.

To address this difficulty, some researchers have begun exploring bug-driven testing paradigms. BugHunter~\cite{chen2025standing} retrieves historical crash knowledge from similar apps to guide GUI testing, but it depends heavily on the availability of high-quality bug data and performs best when relevant apps exist. Other researchers construct knowledge graphs from historical bug reports of the same app~\cite{su2024enhancing,su2022constructing,wang2024knowledge}, encoding past failures into structured knowledge to generate new test cases. However, they require significant manual effort and are limited to a single app, restricting their generalizability.
In this work, our goal is to design a proactive testing approach that leverages code-level patterns as guidance signals, helping exploration prioritize paths that are more likely to expose crashes. 
Although static analysis~\cite{arzt2014flowdroid,qiu2018analyzing,mahmud2022acid,dilhara2018automated,wu2021program,samhi2022difuzer,li2016droidra,yang2015static,li2015iccta,pauck2021jicer,li2022cross,samhi2021raicc,karakaya2021sootfx} has been studied for detecting potential bugs, it faces fundamental challenges in rule design. If the rules are strict, many bug patterns are missed because they do not always deterministically lead to bugs. If the rules are relaxed, the tools generate an overwhelming number of warnings, making the results impractical for manual inspection.
Therefore, rather than treating code analysis as a bug detector, we use it as a provider of exploration guidance signals.
By mapping these suspicious code signals to GUI widgets and execution paths, our approach allows dynamic exploration to prioritize suspicious contexts first, while still maintaining balanced coverage.

We propose {\tool}, an automated GUI testing approach that uses code signals to guide dynamic exploration and prioritize bug detection. {\tool} consists of three modules. {\tool} first collects suspicious code-level signals. We design a set of heuristic rules derived from our empirical study of real-world crashes, which capture representative suspicious patterns and associate them with corresponding GUI widgets. By further backtracking event handlers and call chains, this module extracts early code-level signals that serve as seeds for subsequent analysis.  
Second, {\tool} links the code-level signals to generate execution paths. {\tool} constructs a global Activity Transition Graph (ATG) and performs rough path generation by linking suspicious widgets to navigation flows. To handle missing or obfuscated attributes, we employ LLM-based reasoning for executable path completion, ensuring that each suspicious interaction is aligned with a feasible cross-activity route.  
Third, {\tool} adopts suspicious paths to guide action generation. It combines global path planning, which prioritizes exploration and reduces redundant traversals, with local path generation to adapt to runtime UI variations. An execution-aware monitor continuously checks whether the suspicious paths are exercised as intended and dynamically adjusts the exploration strategy.

We evaluate the effectiveness of {\tool} on 70 crash bugs collected from 42 open-source apps. Compared with 15 widely used and state-of-the-art automated GUI testing baselines, {\tool} achieves the highest bug detection performance, with a recall of 77\%, outperforming the best baseline by 28\%. The activity and code coverage of {\tool} are also comparable to or higher than those of the baselines. To further understand the role of each sub-module, we conduct ablation experiments that confirm the necessity and effectiveness of our design. We also evaluate the usefulness of {\tool} by testing unseen apps from Google Play. Among 116 popular apps, {\tool} successfully detects 21 new crash bugs, of which 15 have already been fixed and 6 have been confirmed by developers.

The contributions of this paper are as follows:
\begin{itemize}
\item We propose {\tool}\footnote{We release the source code, dataset, and experimental results on our website \url{https://github.com/TheMystery123/TraceDroid}. \label{github}}, a novel automated GUI testing approach that proactively explores suspicious paths by combining code-level signals with dynamic exploration. It collects suspicious code-level signals, generates suspicious paths through ATG construction and path completion, and guides dynamic execution to efficiently detect bugs. (Section~\ref{sec_approach})

\item We conduct an empirical study on crash bugs and analyzed their distribution. We also design heuristic rules for different types for future research.

\item Effectiveness and usefulness evaluation of the {\tool} in widely used open-source apps on Google Play with practical bugs detected and confirmed (Section \ref{sec_results}).
\end{itemize}

\section{Motivational Study}
\label{sec_motivation}

\subsection{Data Collection}
\label{sec_motivation_data_collection}
The experimental dataset is collected from GitHub, which has a great number of crash issue reports. We randomly select 300 apps with 5783 issue reports. Two filtering criteria are applied: (1) whether the report describes a crash bug, and (2) whether the report can be mapped to specific code modifications in the linked commit. Two authors manually check the selected apps and issue reports to ensure they have crash bugs. The third author rechecks the results until a consensus is reached. A total of 1639 crash bugs with bug reports are collected. The filtered dataset is then used for manual inspection and classification of crash root causes.

\subsection{Categorizing Crash Bugs}
During the manual label process, we observed that crash bugs in mobile applications exhibit recurring patterns across different projects. Categorizing these patterns can help us design heuristic rules and guide dynamic exploration more effectively. According to the card sorting method~\cite{spencer2009card}, we classify the crash bugs into five major categories, each containing several subcategories that reflect common triggering contexts and functional scenarios. The detailed information is as follows.

\begin{figure*}[htb]
\centering
\vspace{-0.1in}
\includegraphics[width=12.3cm]{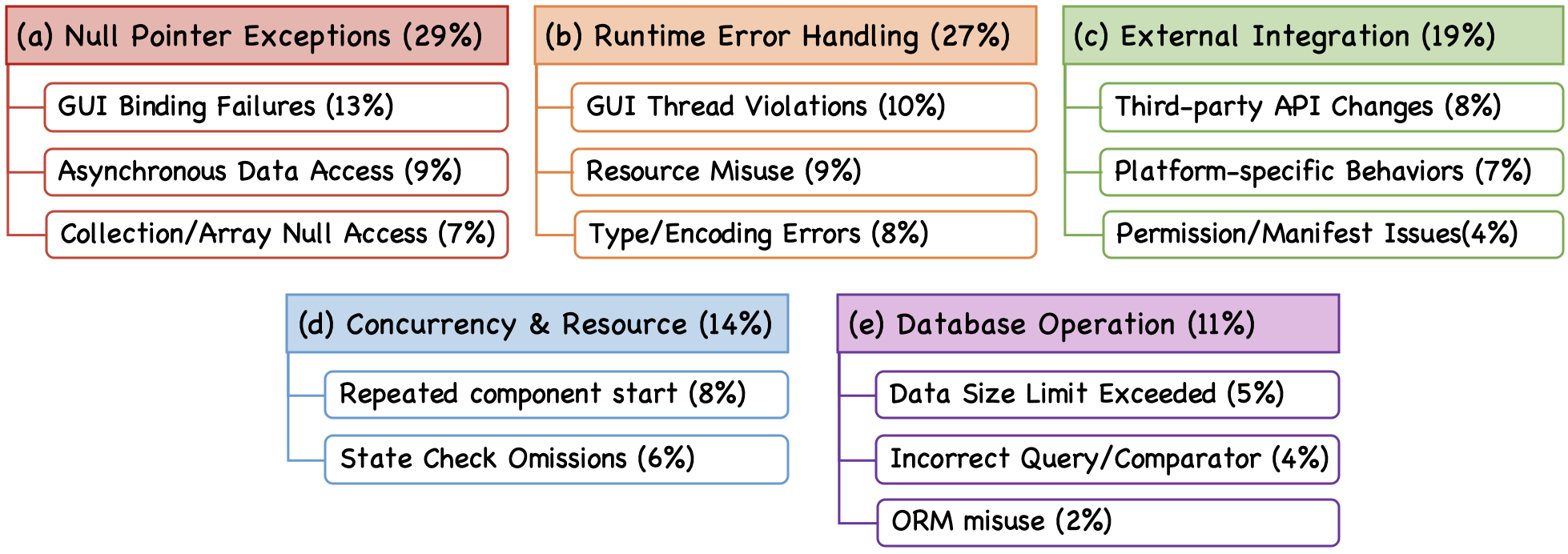}
\vspace{-0.1in}
\caption{Example of 5 categories of crash bug}
\label{fig:pilot-study}
\vspace{-0.2in}
\end{figure*}

\textbf{Null Pointer Exceptions (29\%)}:  
As shown in Fig.~\ref{fig:pilot-study} (a), this category refers to crashes caused by accessing null objects without proper validation. It includes \textit{GUI binding failures}, where views are accessed before data loading is complete; \textit{asynchronous data access}, where callbacks reference uninitialized or released objects; and \textit{collection/array null access}, where data structures are accessed without null checks. These issues frequently occur in functionalities involving data loading and UI updates, such as news feeds, media galleries, or chat lists.

\textbf{Runtime Exceptions and Error Handling (27\%)}:  
As shown in Fig.~\ref{fig:pilot-study} (b), this category covers crashes triggered by missing or insufficient exception handling during critical operations. Subcategories include \textit{UI thread violations}, where UI elements are updated from non-main threads; \textit{resource misuse}, such as invalid resource IDs or missing assets; and \textit{type/encoding errors}, including type casting failures or character encoding mismatches. These bugs are common in background operations like uploads, scans, or data backups.

\textbf{External Integration and Compatibility (19\%)}:  
As shown in Fig.~\ref{fig:pilot-study} (c), this category includes crashes resulting from incompatibilities with external APIs, third-party libraries, or platform-specific behaviors. Subcategories include \textit{third-party API changes} that are not adapted in time; \textit{platform-specific behaviors} that differ across Android versions; and \textit{permission/manifest configuration issues}, such as incorrect permission declarations or forced removal of system-level permissions. These crashes often occur in features that depend on external services or hardware integrations.

\textbf{Concurrency and Resource Management (14\%)}:  
As shown in Fig.~\ref{fig:pilot-study} (d), this category refers to crashes caused by improper handling of concurrent processes or system resources. It includes \textit{repeated component start}, where services or activities are started multiple times in quick succession; \textit{state check omissions}, where component states are not verified before operations; and \textit{non-atomic variable access}, leading to inconsistent states in concurrent scenarios. Such crashes often appear in functionalities involving shared resources, such as content sharing or user status updates.

\textbf{Database Operation Issues (11\%)}:  
As shown in Fig.~\ref{fig:pilot-study} (e), this category involves crashes related to database misuse or exceeding system limitations. It includes \textit{data size limit exceeded}, such as storing large BLOBs beyond SQLite constraints; \textit{incorrect query/comparator}, where violations of comparator contracts cause crashes during sorting; and \textit{ORM misuse}, where APIs from frameworks like Room are incorrectly applied. These issues are prevalent in data-heavy applications such as note-taking or accounting software.

\section{Approach}
\label{sec_approach}
We propose {\tool}, an automated GUI testing approach designed to proactively explore abnormal execution contexts by combining code-level signals with dynamic exploration. The core idea is that although GUI testing is the practical way to detect crashes, unguided exploration often fails to achieve the precise interaction sequence that triggers crashes. {\tool} utilizes lightweight code analysis not to directly detect crashes, but to identify suspicious code-level signals that may indicate unstable behavior and use them as guidance for GUI exploration.

\begin{figure*}[htb]
\centering
\includegraphics[width=13.7cm]{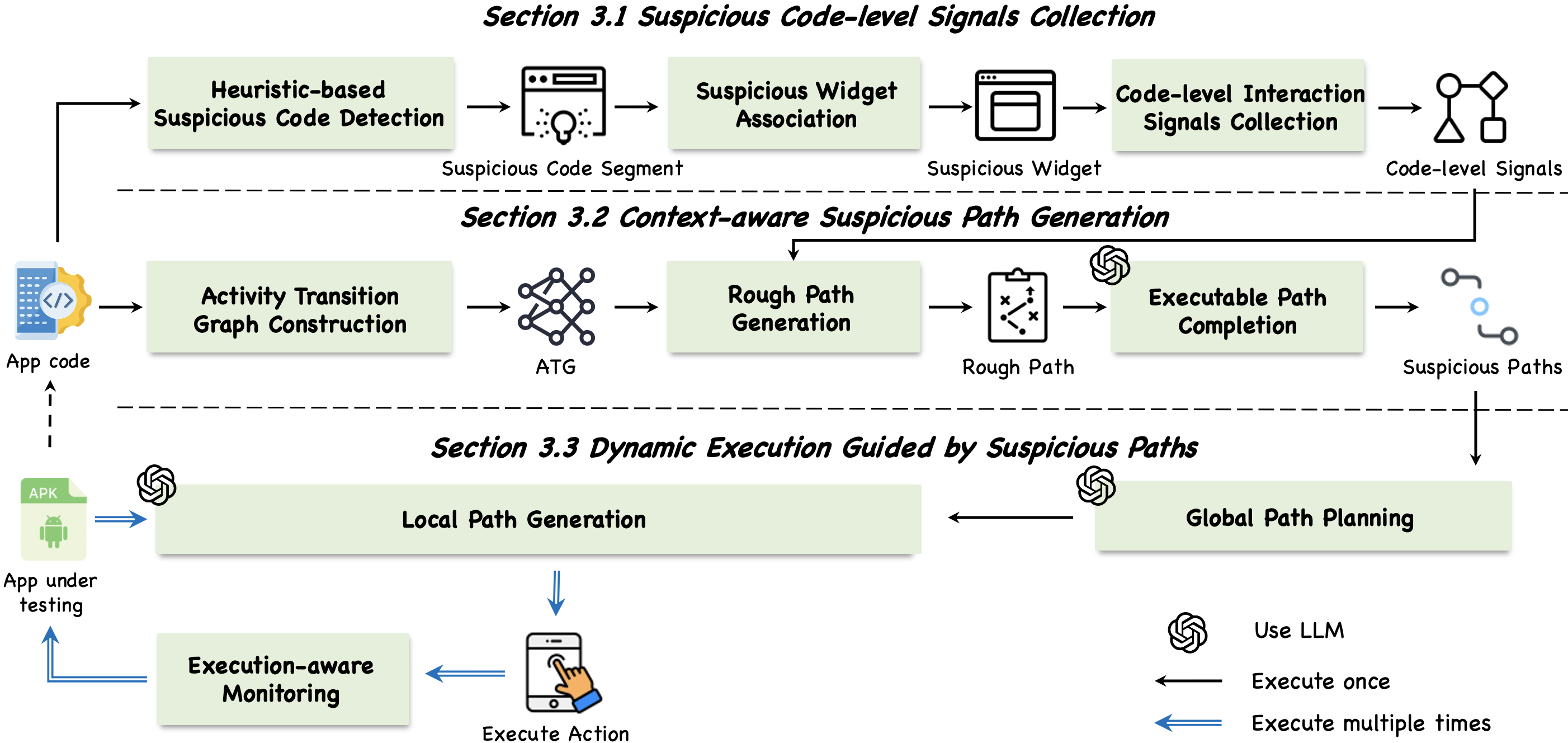}
\vspace{-0.1in}
\caption{Overview of {\tool}}

\label{fig:overview}
\vspace{-0.1in}
\end{figure*}

As shown in Fig.~\ref{fig:overview}, {\tool} consists of three main modules: \textit{Suspicious Code-level Signals Collection}, \textit{Context-aware Suspicious Path Generation}, and \textit{Dynamic Execution Guided by Suspicious Paths}. The first module applies heuristic rules to detect suspicious code patterns, associates them with corresponding GUI widgets, and collects their code-level interaction signals. The second module constructs the Activity Transition Graph (ATG) of the target app, generates rough paths, and employs LLM-based reasoning to complete executable paths, producing a set of suspicious paths. The third module conducts dynamic execution guided by these suspicious paths, integrating global path planning with local path generation, while an execution-aware monitoring mechanism validates execution and records crashes. 
Meanwhile, the suspicious paths discovered by {\tool} can also be integrated into existing LLM-based GUI testing approaches as prompt-level hints, potentially enhancing their bug detection ability.

\subsection{Suspicious Code-Level Signals Collection}
\label{subsec_suspicious_segment_localization}
The purpose of this module is to generate inspiration for GUI testing by tracing from suspicious code fragments to their corresponding GUI widgets and collecting their \textit{code-level interaction signals}. Based on the empirical study in Section~\ref{sec_motivation}, we design heuristic rules that capture five representative categories of crash-inducing patterns. Applying these rules to the app's source code and configuration files produces a set of suspicious code segments. Each segment is then associated with candidate GUI widgets by analyzing layout files and event-binding relations. Finally, we extract the interaction signals from method calls and callback chains that are linked to these widgets. 
The output of this module is a set of \textit{suspicious code-level signals}, each consisting of a suspicious code fragment, the associated GUI widget, and its triggering signals. These signals serve as seeds for the path generation module in Section~\ref{subsec_path_mapping}, where they are further aligned with execution-level paths.

\begin{figure*}[htb]
\centering
\vspace{-0.1in}
\includegraphics[width=13.7cm]{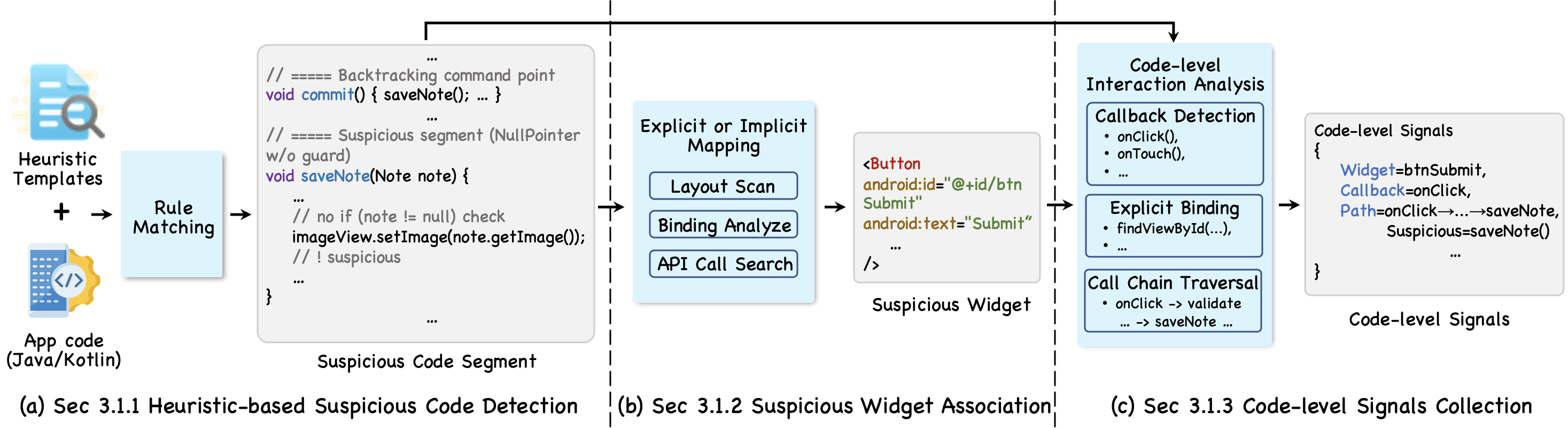}
\vspace{-0.1in}
\caption{Example of suspicious code-level signals collection}
\label{fig:static}
\vspace{-0.1in}
\end{figure*}

\subsubsection{Heuristic-Based Suspicious Code Detection}
\label{subsubsec_code_detection}
The first step of our approach is to automatically locate code segments as shown in Fig. \ref{fig:static} (a), that exhibit suspicious patterns. Based on the findings of our motivational study in Section~\ref{sec_motivation}, we designed a set of heuristic detection rules that capture five representative categories of crash-inducing patterns: \textit{NullPointer}, \textit{RuntimeError}, \textit{ExternalIntegration}, \textit{Concurrency}, and \textit{DBOperation}. In total, we implemented 73 fine-grained rules, with 10-18 sub-rules per category, together with several variants to account for implementation differences across projects. The detailed list of rules is made publicly available in our GitHub repository\textsuperscript{\ref{github}} for reproducibility.

The rules are designed by following two principles. First, each rule encodes empirical patterns distilled from real-world bug reports, ensuring that the rule set is grounded in practical evidence. Second, rules consider \textit{contextual conditions} rather than isolated statements. For instance, dereference operations (e.g., \texttt{obj.method()} or \texttt{obj.field}) are only flagged as suspicious if they appear along a control-flow path where no null-check (e.g., \texttt{if (obj != null)}) dominates the operation. Similarly, a call to \texttt{setText()} is only flagged when it is reachable from a background thread without a dispatcher to the UI thread, indicating a potential runtime violation. This context-sensitive design distinguishes our rules from simple syntactic pattern matching and reduces false positives.  
Across all rules, a single rule involves between one and three contextual conditions (average of 1.9 conditions per rule). For \textit{NullPointer} crashes (16 rules), where contextual conditions mainly capture missing null-dominance checks or premature dereference of objects before proper initialization. For \textit{RuntimeError} crashes (18 rules),  their contextual conditions frequently encode whether UI-modifying operations are reachable from background threads, or whether lifecycle callbacks are invoked without required state guards. For \textit{ExternalIntegration} crashes (14 rules), whose contextual conditions focus on platform and environment dependencies, such as missing permission checks, unavailable external services, or incompatible API usage. For \textit{Concurrency} crashes (13 rules), whose contextual conditions emphasize temporal and state-related properties, including repeated lifecycle invocations, missing synchronization, or reuse of stale component states. For \textit{DBOperation} crashes (12 rules), where contextual conditions mainly describe database-specific execution constraints, such as unclosed cursors, incomplete transactions, or invalid access after database closure.

To implement the detection, we parse all Java/Kotlin source files into Abstract Syntax Trees (ASTs) and traverse the control-flow and data-flow graphs. Each rule is applied to the AST representation of methods and callback handlers, scanning not only the target operation but also the surrounding conditions and preceding statements. Layout files and resource identifiers (e.g., \texttt{R.id.*}) are also parsed to enrich the analysis, since many suspicious behaviors are tied to specific widgets or event handlers. For concurrency-related rules, the analysis further considers the temporal proximity of repeated lifecycle calls (e.g., consecutive \texttt{startService()} invocations) and checks whether the corresponding termination or state validation is missing. The output of this step is a set of suspicious code segments annotated with their rule category and location. 

\subsubsection{Suspicious Widget Association}
\label{subsubsec_component_mapping}
After suspicious code segments are detected, the next step is to associate them with the corresponding GUI widgets.
We design a multi-dimensional matching strategy that integrates information from three sources: layout XML files, event binding relations, and API call semantics. This strategy ensures that suspicious segments are mapped to the correct widgets even when the connections are indirect, implicit, or spread across multiple files. 
As shown in Fig. \ref{fig:static} (b), our implementation covers both explicit mappings (e.g., direct references to \texttt{R.id.*}) and implicit mappings (e.g., dynamically inflated layouts or inherited listeners). 

First, we parse layout XML files that are connected to the suspicious segment. 
We scan for layout inflation calls, including \texttt{setContentView(R.layout.*)} and \texttt{inflate(R.layout.*, ...)}. For each relevant layout, we extract all widgets and record their attributes: (1) the component ID from the \texttt{android:id} attribute, (2) the widget type from the XML tag name (e.g., \texttt{Button}), (3) textual information from the \texttt{android:text} field or linked string resources, and (4) the relative position in the XML hierarchy, which preserves parent-child relationships. 

Second, we analyze event binding information. Suspicious segments often occur inside event listeners, so we track calls that register event handlers, including setOnClickListener(), setOnLongClickListener(), setOnTouchListener(), etc. We further resolve indirect bindings where the listener registration is delegated through helper methods or abstract classes. These bindings are linked to widget IDs by analyzing the parameters of registration calls and their surrounding control-flow scope. 

Third, we incorporate API call semantics. For example, if a suspicious segment invokes \texttt{setText()} on a \texttt{TextView}, the mapped widget should be of type \texttt{TextView} or its subclass. This semantic verification ensures type-level consistency between suspicious operations and candidate widgets.

\subsubsection{Code-Level Interaction Signals Collection}
\label{subsubsec_approach_RiskyCodeToGUI} 
Once suspicious code segments are associated with candidate widgets, the next step is to collect code-level interaction signals that indicate how these widgets may trigger the suspicious operations. As shown in Fig.~\ref{fig:static} (c), we extract key interaction-related signals from the code that can later serve as seeds for suspicious path generation.

We begin by parsing the method that contains the suspicious code into its Abstract Syntax Tree (AST) and control-flow graph (CFG). The backward traversal searches for enclosing or upstream methods that are recognized as event callbacks, which include framework-defined callbacks (\texttt{onClick()}, \texttt{onItemSelected()}, \texttt{onTouch()}, \texttt{onLongClick()}), adapter callbacks (\texttt{getView()}), and system lifecycle methods (\texttt{onActivityResult()}, \texttt{onResume()}). When a suspicious statement is located inside such a callback, we infer that the corresponding widget interaction can directly trigger it.

If the connection is indirect, we extend the analysis to cover two additional scenarios. First, we resolve \textit{explicit bindings}, where event listeners are dynamically registered in the code. We analyze calls such as \texttt{findViewById()} followed by \texttt{setOnClickListener()} (or variants like \texttt{setOnTouchListener()}, \texttt{addTextChangedListener()}). The integer parameter of \texttt{findViewById()} is traced back to a resource ID defined in the layout XML, which is then matched to a widget entry. Second, we address \textit{implicit bindings}, where handler methods are invoked through indirection, such as XML-declared \texttt{android:onClick} attributes or reflection-based invocations. For these cases, we scan layout files for attributes that reference method names and match them to methods in the AST. Reflection calls are resolved conservatively by analyzing string literals passed to \texttt{Class.getMethod()} or \texttt{Method.invoke()}, recording all feasible method targets. 
The backtracking also considers multi-step call chains. For example, a suspicious segment in \texttt{saveNote()} may be invoked by \texttt{onClick()}, but only after passing through helper methods (e.g., \texttt{validate()} $\rightarrow$ \texttt{commit()} $\rightarrow$ \texttt{saveNote()}). We traverse the call graph until reaching a GUI entry point (a callback registered to a widget or lifecycle). This traversal is bounded to avoid infinite loops, and recursive calls are resolved by summarizing the call site once.

The final output of this step is a set of structured \textit{code-level interaction signals}, consisting of: (1) the widget ID, type, and layout file in which it is defined, (2) the sequence of event-handling methods leading to the suspicious segment, and (3) the relative position of the suspicious code within this chain. These signals are used as seeds for the suspicious path generation module in Section~\ref{subsec_path_mapping}, where they are aligned with executable paths of the app.

\begin{figure*}[htb]
\centering
\includegraphics[width=13.7cm]{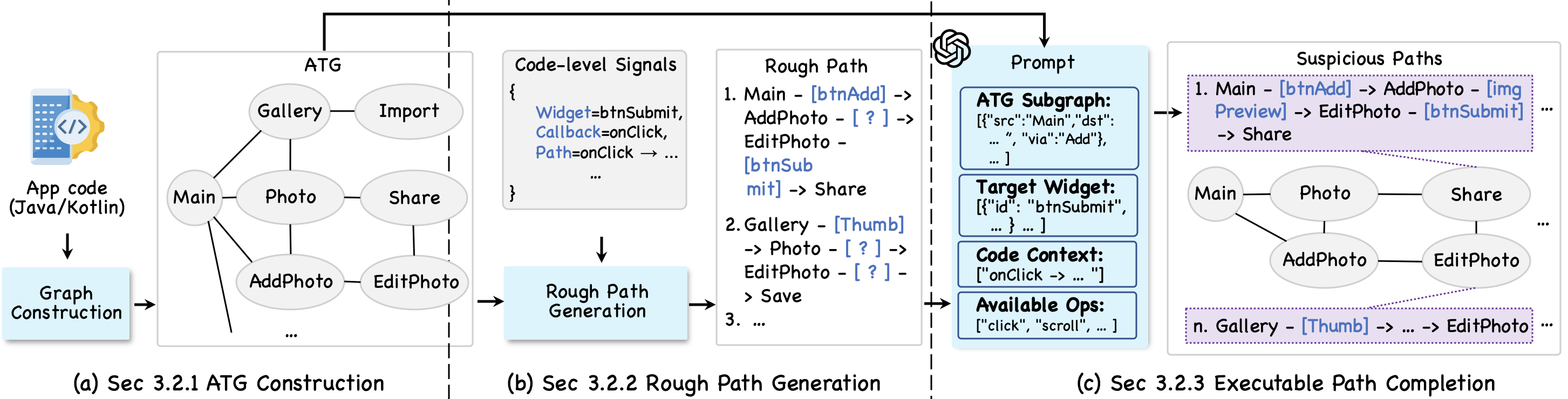}
\vspace{-0.1in}
\caption{Example of context-aware suspicious path generation}
\label{fig:mapping}
\vspace{-0.1in}
\end{figure*}

\subsection{Context-Aware Suspicious Path Generation}
\label{subsec_path_mapping}
After obtaining code-level interaction signals from the first module, the second module aims to map these signals to concrete GUI widgets and align them with executable navigation paths. The key challenge here is that static analysis often provides only partial information, and the same widget may appear in multiple activities or layouts. 
To address this, {\tool} first performs \textit{Activity Transition Graph (ATG) Construction} by analyzing both manifest declarations and navigation code, ensuring that all possible inter-activity transitions are captured. 
It then conducts \textit{Rough Path Generation}, which leverages identifiers, textual properties, type compatibility, and binding relations to associate code-level signals with candidate widgets and approximate their navigation paths within the ATG. 
The output of this module is a set of \textit{suspicious paths} that connect code-level signals with their corresponding GUI widgets and cross-activity navigation traces, serving as direct guidance for the dynamic execution module in Section~\ref{subsec_dynamic_execution}.

\subsubsection{Activity Transition Graph Construction}
\label{subsubsec_ATG_construction}
The first step of this module is to construct a global Activity Transition Graph (ATG) that captures all feasible navigation flows within the app. We begin by parsing the \textit{AndroidManifest.xml} file to enumerate all declared activities and services. For each activity, we record its name, intent filters (including actions, categories, and data types), and launch modes. This step establishes the node set of the ATG as shown in Fig. \ref{fig:mapping} (a).

Next, we analyze the app's source code to extract inter-activity transitions. Explicit transitions are identified by locating invocations of APIs such as \textit{startActivity()}, \textit{startActivityForResult()}, and \textit{startService()}. When the \textit{Intent} instantiation explicitly specifies a target class (e.g., \textit{new Intent(this, TargetActivity.class)}), we directly add a directed edge from the caller activity to \textit{TargetActivity}. We also account for indirect intent construction, such as when the target component is passed as a variable or obtained from a helper method, by propagating symbolic references along the control-flow graph until the target class is resolved.

Implicit transitions are extracted by cross-referencing manifest-declared intent filters with code-level intent usage. For example, when an activity declares an intent filter with action \textit{android.intent.action.VIEW} and category \textit{android.intent.category.DEFAULT}, any \textit{sendBroadcast()} or \textit{startActivity()} call with a matching action-category pair is linked to that activity. This matching process allows us to capture navigation edges that are not explicitly defined in the code but are nevertheless valid at runtime.

In addition to activity-to-activity transitions, we also incorporate fragment-based navigation. We detect fragment transactions (e.g., \textit{FragmentTransaction.add()}, \textit{replace()}, \textit{show()}) and model fragments as intermediate nodes within the ATG. Edges are added from the host activity to its fragments, and between fragments, when a transaction replaces one with another. This extension allows the ATG to capture intra-activity navigation that is critical in modern Android applications.

Unlike traditional ATG construction that only records activity-level edges, our approach further annotates each transition with the widgets or triggers discovered in static analysis. For instance, if an invocation of \texttt{startActivity()} is guarded by a button click, the corresponding edge is annotated with the widget's ID and type (e.g., \texttt{btnAdd}, \texttt{ImagePreview}). In cases where partial information is missing, such as when no \texttt{id} or \texttt{text} can be resolved, the annotation is left blank. Similarly, when static analysis reveals an actionable widget but fails to link it to a specific transition edge, we conservatively attach the widget to the activity node itself.

\subsubsection{Rough Path Generation}
\label{subsubsec_approach_MultiDimMatching}
After obtaining the enriched ATG with annotated but often incomplete widget information, the next step is to generate \textit{rough paths} that connect code-level signals to candidate navigation traces. Some edges are annotated with precise identifiers (e.g., \texttt{R.id.btnAdd}), while others may only reveal a callback name or widget type, and certain components remain entirely unlinked. To bridge these gaps, we design a matching strategy that systematically integrates both code-level and UI-level evidence as shown in Fig. \ref{fig:mapping} (b).

Specifically, for each code-level signal ${widget,callback,path}$ obtained in Module~\ref{subsubsec_approach_RiskyCodeToGUI}, we attempt to match it with candidate widgets discovered in the ATG.  Some are extracted from static code (IDs, listener registrations), others from resource files or runtime hierarchy dumps (text labels, widget types, structural positions). To tolerate incompleteness, we use a scoring across four dimensions:
(1) \textit{ID consistency.}. When an explicit ID is available, we normalize it (removing \texttt{R.id.} prefixes and aliases) and directly compare.
(2) \textit{Textual similarity.} When a label or content description is present, we compute string similarity between code constants and resource texts. For instance, a suspicious call \texttt{setText(``Submit'')} can be aligned with a widget in XML defined as \texttt{android:text=``Submit''}.
(3) \textit{Type compatibility.} The widget type implied by the API usage (e.g., \texttt{setImage()} $\Rightarrow$ \texttt{ImageView}) is cross-checked with the XML or hierarchy type.
(4) \textit{Callback linkage.} If the suspicious code resides in a listener (e.g., \texttt{onClick()}), we check whether the candidate widget is registered with the same listener type.

A candidate widget is accepted if the combined score passes the threshold of 0.8. When multiple candidates qualify, we keep all of them for later disambiguation. Through this process, the ATG edges that were partially annotated become enriched with concrete or candidate widgets, forming \textit{rough paths} that connect suspicious code-level signals to navigation traces. These rough paths still contain missing slots (e.g., ``\texttt{[?]}'' as shown in Fig. \ref{fig:mapping} (b)),
where LLM-based reasoning will complete and refine them into executable suspicious paths.

\subsubsection{Executable Path Completion}
\label{subsubsec_context_path_alignment}
After obtaining rough paths from the previous stage, some edges are annotated with missing IDs, some widgets lack textual labels, and some navigations are unresolved. To turn these rough traces into executable navigation sequences, we employ an LLM to perform \textit{executable path completion}. The model leverages both structural information from the ATG and contextual hints from the code-level signals to fill in the missing steps and produce realistic interaction paths.

For each rough path, we serialize a structured prompt that contains the following fields as shown in Fig. \ref{fig:mapping} (c). (1) ATG Subgraph. A two-hop neighborhood around the source and target activities, represented as a list of candidate transitions (e.g., \texttt{[{"src": "Main", "dst": "AddPhoto", "via": "Button:id=btnAdd"}, ... ]}). (2) Target Widget. The suspicious widget to be triggered, with any available attributes such as \texttt{id}, \texttt{text}, or \texttt{type}. (3) Code-level Signals. The callback chain and local guards obtained from backtracking (e.g., \texttt{onClick -> saveNote()}, with required conditions such as \texttt{note != null}). (4) Available Operations. The primitive actions executable on device (\texttt{click}, \texttt{scroll}, \texttt{input}, etc.) and selectors supported (\texttt{id}, \texttt{text}, \texttt{xpath}).

The LLM then outputs a candidate action sequence in JSON format, where each step specifies the \texttt{activity}, \texttt{selector}, \texttt{operation}, and optional \texttt{params}. To ensure reliability, we validate the generated paths using two lightweight checks: (1) ATG consistency: all inter-activity hops must exist in the subgraph or correspond to valid implicit intents. (2) Widget existence: each selector must correspond to an actual widget in the layout or view hierarchy dump. Only validated plans are retained as \textit{suspicious paths}. Paths that share long prefixes are merged, allowing reuse across multiple suspicious contexts.

\subsection{Dynamic Execution Guided by Suspicious Paths}
\label{subsec_dynamic_execution}
This module uses the suspicious paths as explicit guidance while still retaining the flexibility to adapt to runtime conditions. The design consists of three complementary steps. First, a \textit{Global Path Planning} phase schedules which suspicious paths should be explored and in what order, balancing high-risk prioritization with coverage considerations. Second, a \textit{Local Path Generation} phase translates each suspicious path into concrete executable actions based on the current GUI state, resolving ambiguities such as multiple candidate widgets or layout variations. Third, an \textit{Execution-aware Monitoring} phase observes runtime feedback, checking whether the intended suspicious path was successfully executed, whether it triggered abnormal behaviors, and whether recovery actions (e.g., restarting the app) are required.

\begin{figure*}[htb]
\centering
\includegraphics[width=13.7cm]{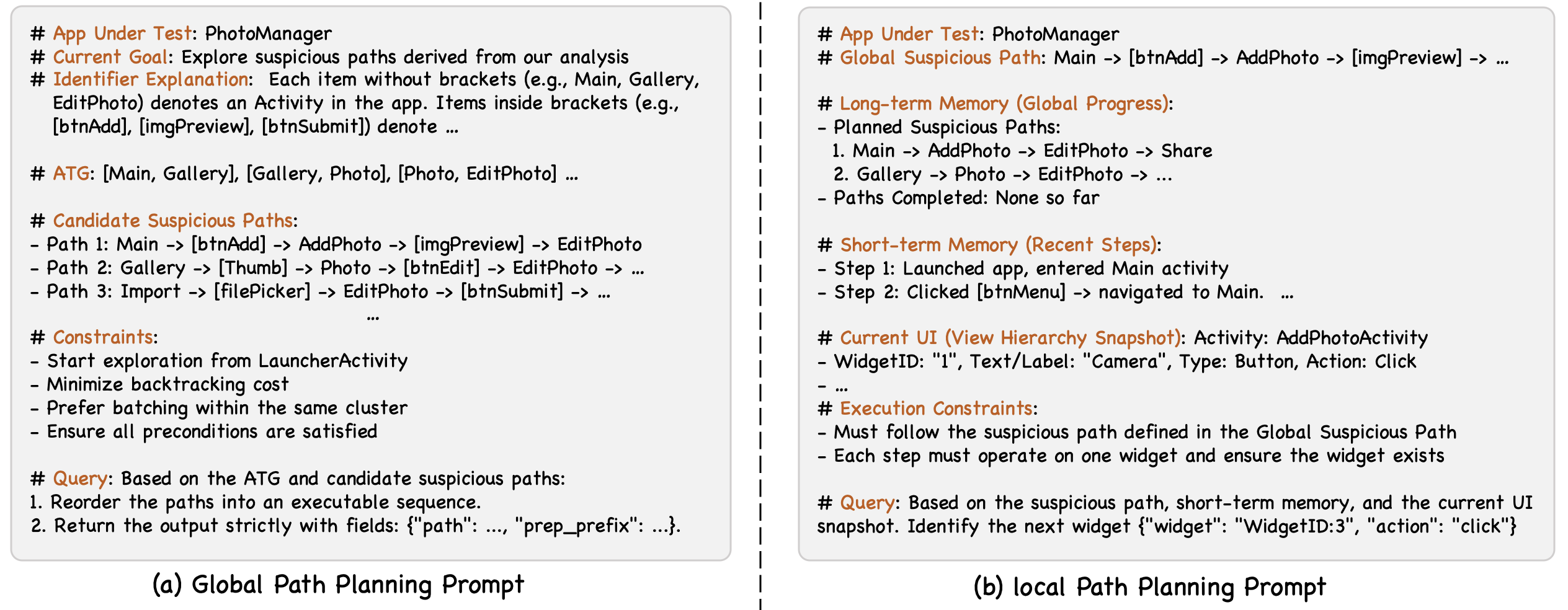}
\vspace{-0.1in}
\caption{Example of prompt in global and local path planning}
\label{fig:prompt}
\vspace{-0.2in}
\end{figure*}

\subsubsection{Global Path Planning}
\label{subsubsec_global_planning}
Before execution starts, {\tool} performs a one-time global scheduling that determines the order in which suspicious paths should be explored. The Activity Transition Graph (ATG) provides the structural backbone, but does not prescribe an execution sequence. This step therefore converts the unordered suspicious paths into an executable itinerary.

To achieve this, we design a structured prompt for the LLM that contains four elements: (1) an identifier explanation that clarifies the meaning of activities and widgets; (2) the extracted ATG edges, which encode all feasible activity transitions; (3) the list of candidate suspicious paths derived from widget mapping and backtracking; and (4) explicit constraints such as starting from the launcher activity, minimizing backtracking cost, batching paths within the same cluster, and ensuring all preconditions are satisfied.

The query at the end of the prompt explicitly instructs the LLM to reorder the candidate paths into an executable sequence and return the output in a strict JSON format. This structured format avoids ambiguity and allows {\tool} to directly consume the LLM's plan. After receiving the output, {\tool} validates each transition against the ATG using graph traversal. If the LLM proposes an unreachable hop, it is replaced by the deterministic shortest path. Missing or malformed fields are also repaired through lightweight deterministic checks. This validation ensures that the final itinerary is both syntactically correct and semantically executable, while the LLM provides high-level ordering that exploits locality and shared prefixes. The result of this step is a deterministic itinerary, where each suspicious path appears exactly once. This itinerary is then passed to the Local Path Generation phase, which translates each path into concrete interaction steps at runtime.

\subsubsection{Local Path Generation}
\label{subsubsec_local_generation}
After global planning produces a high-level itinerary, {\tool} must translate each scheduled suspicious path into concrete interaction steps that can be directly executed on the device. Since the planned paths are abstract and runtime UIs may contain multiple candidates or dynamic variations, local path generation provides step-by-step grounding with explicit memory and context management.

Each generation task is formulated as a structured prompt to the LLM. 
(1) Global Suspicious Path selected from global planning, which ensures the LLM operates within the intended navigation trajectory.
(2) Long-term Memory (Global Progress) is the list of planned suspicious paths and their current execution status. This avoids revisiting already completed branches and ensures continuity across multiple paths. (3) Short-term Memory (Recent Steps) means the last few actions executed, which provides local execution context and prevents loops or redundant actions. (4) Current UI (View Hierarchy Snapshot) means the annotated GUI screenshot, including widget IDs, labels, types, and clickable/input attributes. These annotations serve as explicit grounding for possible actions.

The query then asks the LLM to generate the next single action in JSON format. By constraining output to one step per query and enforcing a fixed schema, {\tool} prevents free-form responses and facilitates automatic replay on the device. Through this design, local path generation combines global trajectory alignment, explicit memory for long-term and short-term contexts, and accurate grounding in the current UI snapshot. Unlike conventional GUI testing that executes pre-computed scripts or random interactions, this approach enables {\tool} to adaptively resolve ambiguities at runtime while ensuring consistency with suspicious paths. 

\subsubsection{Execution-Aware Monitoring}
\label{subsubsec_execution_monitoring}
The final step of dynamic execution is to monitor whether each suspicious path has been faithfully exercised and whether its execution exposes abnormal behaviors. Unlike conventional crash detection that relies solely on coarse-grained system logs, {\tool} integrates runtime monitoring with both path context and UI state to provide fine-grained validation and robust recovery.

The monitoring process takes as input. (1) Scheduled Suspicious Path. The planned sequence of activities and widgets that should be executed.
(2) Runtime Execution Trace. The sequence of actions actually performed, each paired with its corresponding view hierarchy snapshot.
(3) System-level Feedback. Signals from \texttt{logcat}, tombstone files, exception stack traces, and process status.

During execution, each action is verified against the suspicious path. If a widget mismatch occurs (e.g., expected \texttt{btnSubmit} is absent due to a dynamic layout), the path is flagged as incomplete, and a fallback strategy continues from the nearest valid state.
{\tool} continuously observes abnormal behaviors across three categories: (1) Crash events. Uncaught exceptions or process terminations, identified by stack traces and exit codes. (2) Freeze or deadlock. Detected through repeated failed UI hierarchy dumps without progress. (3) Path deviations. Divergence between the intended suspicious path and the executed trace.
After each path terminates, the monitor ensures the app is reset to a clean state. Crashed apps are relaunched automatically, while incomplete paths are retained for possible re-execution if resources allow.

\subsection{Implementation}
In our implementation, we utilize the GPT-4 Vision model\footnote{\url{https://platform.openai.com/docs/models/gpt-4o}} (gpt-4o) as MLLM to handle both textual and visual information during the testing process. It obtains the view hierarchy file of the current GUI page through UIAutomator~\cite{uiautomator} to extract text information of the input widgets.
{\tool} is implemented as a fully automated GUI testing tool, which uses or extends the following tools: VirtualBox and the Python pyvbox~\cite{pyvbox} for running and Android Debug Bridge (ADB)~\cite{Adb} for interacting with the app.

\section{Experiment Design}
\label{sec_Experiment_Design}
\begin{itemize}[leftmargin=*]
\item \textbf{RQ1: (Detection Effectiveness and Coverage)}
How effective is {\tool} in exposing crash behaviors and achieving testing coverage?
\end{itemize}

For RQ1, we evaluate the effectiveness of {\tool} in detecting crash bugs and compare it with widely used and state-of-the-art automated GUI testing baselines.

\begin{itemize}[leftmargin=*]
\item \textbf{RQ2: (Ablation Study)}
What is the contribution of each (sub-)module of {\tool} to the overall detection effectiveness and coverage?
\end{itemize}

For RQ2, we conduct ablation experiments by selectively disabling modules of {\tool} to evaluate their impact on detection rate and coverage.

\begin{itemize}[leftmargin=*]
\item \textbf{RQ3: (Usefulness Evaluation)}
How useful is {\tool} in real-world scenarios?
\end{itemize}

For RQ3, we use {\tool} to automatically explore the app from Google Play, detect unseen bugs, and issue the detected bugs to the development team.

\subsection{Dataset}
\label{subsec_experiment_dataset}
\textbf{For RQ1 and RQ2,} the experimental dataset is derived from the same collection process described in our pilot study in Section~\ref{sec_motivation}, but using a disjoint set of apps. In detail, we randomly selected 150 different apps with 183 crash bug reports that do not overlap with those 300 in the pilot study, following the same app names to guarantee disjointness. We also keep the ones with at least one update after Jan. 2024 in 15 App categories.

Then, we manually reviewed all collected issue reports and filtered apps based on the following criteria:
(1) All baseline tools can run on the app.
(2) UIAutomator~\cite{uiautomator} can extract both the view hierarchy file and GUI screenshots.
(3) The crash bugs can be reliably reproduced in the given app version.
Finally, 42 apps with 70 crash bugs remain as the evaluation dataset for effectiveness and ablation studies.  The distribution of these bugs across categories (\textit{NullPointer}, \textit{RuntimeError}, \textit{ExternalIntegration}, \textit{Concurrency}, and \textit{DBOperation}) follows the proportions observed in the pilot study, ensuring that the evaluation dataset reflects real-world crash characteristics.

To further characterize the complexity of the evaluated crashes, we manually inspected the reproduction steps of all 70 crash bugs. We find that the crash-triggering paths require between 4 and 16 GUI interactions, with an average of 9.1 steps.
These interactions typically involve a mixture of navigation actions, form inputs, and repeated backtracking across multiple activities, rather than single-step or immediately failing operations. This confirms that the evaluated crashes are generally non-trivial and require multi-step interaction sequences to be triggered.

\textbf{For RQ3,} we further evaluate the usefulness of {\tool} on real-world applications from Google Play.
We randomly selected 200 popular apps across different categories, ensuring diversity in functionality and download volume. To ensure reproducibility and fair comparison, we only retain apps that satisfy the following constraints:
(1) The apps can be successfully decompiled and their source code retrieved, so that our heuristic rules can be applied.
(2) UIAutomator can obtain the view hierarchy and screenshots during testing.
After filtering, we obtained a set of 116 apps as the evaluation target. For each detected crash bug, we created a detailed issue report including reproduction steps and submitted it to the corresponding developers through issue trackers or email. Developer confirmations and fixes are used as ground truth to assess usefulness.

\subsection{Baselines}
\label{subsec_experiment_baseline}
To demonstrate the advantage of {\tool}, we compare it with 15 commonly used and state-of-the-art automated GUI testing techniques.
We roughly divide them into random-/rule-based, model-based, and learning-based methods, to facilitate understanding.
For random-/rule-based methods, we use Monkey~\cite{Monkey} and Droidbot~\cite{li2017droidbot}, Time-Machine~\cite{dong2020time}. For model-based methods, we use Ape~\cite{gu2019practical}, Fastbot~\cite{cai2020fastbot}, ComboDroid~\cite{wang2020combodroid}. For learning-based methods, we use Humanoid~\cite{li2019humanoid} and Q-testing~\cite{pan2020reinforcement}.
For LLM-based methods, we use AppAgent~\cite{yang2023appagent}, GPTDroid~\cite{liu2024make}, DroidAgent~\cite{yoon2023autonomous}, AUITestAgent~\cite{hu2024auitestagent}, VisionDroid~\cite{liu2024vision}, and Guardian~\cite{ran2024guardian}.
For the bug-driven method, we use Bughunter~\cite{chen2025standing}, a bug-aware automated GUI testing approach that generates exploration paths guided by bug reports from similar apps, utilizing a combination of LLMs and RAG.

\subsection{Experimental Setup}
\label{subsec_experiment_Setup}
We implement {\tool} based on the gpt-4o API which is released on the OpenAI website. We deploy the baselines and {\tool} on a 64-bit Ubuntu 18.04 machine and evaluate them on Google Android 7.1 emulators.
Following common practice~\cite{gu2019practical,li2017droidbot}, we register separate accounts for each app that requires login and write the login scripts, and during testing reset the account data before each run to avoid possible interference. To ensure fair and reasonable use of resources, we set up the running time of each tool in one app to 60 minutes, which is widely used in other GUI testing studies~\cite{su2021benchmarking,fan2018large,li2017droidbot,gu2019practical}.
We manually optimized the automated GUI testing tools based on the configuration suggestions provided by the method and applied them.
We run each tool three times and obtain the average performance to mitigate potential bias.

\begin{table}[t]
\centering
\renewcommand\arraystretch{0.9} 
\footnotesize
\vspace{-0.1in}
\caption{Per-category bug detection recall in RQ1.}
\vspace{-0.1in}
\label{tab:rq1_bug_recall}
\begin{tabular}{lcccccc}
\toprule
\textbf{Metric} & \textbf{NullPointer} & \textbf{RuntimeError} & \textbf{ExternalIntegration} & \textbf{Concurrency} & \textbf{DBOperation} & \textbf{Overall}\\
\midrule
Recall & 0.81 & 0.83 & 0.71 & 0.70 & 0.71 & 0.77\\
F1-score & 0.89 & 0.91 & 0.83 & 0.82 & 0.83 & 0.87\\
\bottomrule
\end{tabular}
\vspace{-0.1in}
\end{table}

\subsection{Evaluation Metrics}
For bug detection, we use the number of bugs detected (\#Bugs) by automated GUI testing tools and the recall rate (Recall) to assess the effectiveness of {\tool} in identifying true bugs. Recall measures the ability of {\tool} to identify all relevant bugs present within the app. To calculate recall, we compare the number of bugs detected by {\tool} with the total number of bugs confirmed by developers in the issue reports.
For testing coverage, we obtain the activity coverage and code coverage~\cite{he2020textexerciser,liu2017automatic,arnatovich2018mobolic,wang2020combodroid,wang2021vet,li2019humanoid}, in which we treat the activities defined in the \textit{AndroidManifest.xml} file of an Android app as the whole set of  activities~\cite{pan2020reinforcement,Android,su2017guided}.
For bug detection time, we record the timestamp at which each crash bug is first triggered during the 60-minute testing window. If an app contains multiple crash bugs, each bug has an associated detection time. We compute the \textit{average bug detection time} by taking the arithmetic mean of all detection times of the bugs that are successfully exposed by a tool within the app. For example, if three crash bugs are detected at 5, 18, and 42 minutes respectively, the average detection time is $(5+18+42)/3=21.7$ minutes. If a bug is not detected within the allotted time, it is excluded from the average since its detection time is undefined.

\section{Result}
\label{sec_results}
\subsection{Performance of Bug Detection (RQ1)}
\label{sec_results_RQ1}

We first analyze the bug detection performance of {\tool} across different crash categories. Among the 54 detected bugs, NullPointer (17), RuntimeError (15), ExternalIntegration (10), Concurrency (7), and DBOperation (5). As shown in Table~\ref{tab:rq1_bug_recall}, {\tool} achieves relatively high recall and F1-score for all five categories, with recall values ranging from 0.70 to 0.83 and F1-scores ranging from 0.82 to 0.91.
Among them, \textit{RuntimeError} and \textit{NullPointer} crashes are detected with the highest effectiveness. In particular, \textit{RuntimeError} achieves the highest F1-score (0.91), followed by \textit{NullPointer} (0.89), indicating that these crash types can be detected with both high recall and stable performance.
This is mainly because crashes in these categories often exhibit relatively explicit code-level patterns and well-defined execution contexts. These patterns can be precisely localized by suspicious-code detection and effectively mapped to triggering GUI interactions, allowing {\tool} to prioritize and reconstruct the corresponding crash-triggering paths.

\begin{table}[t]
\renewcommand\arraystretch{0.9} 
\caption{Performance of {\tool} and baselines}
\vspace{-0.05in}
\label{tab:RQ1}
\centering
\footnotesize
\begin{tabular}{p{3.1cm}<{\centering} | p{1.5cm}<{\centering} | p{1.5cm}<{\centering} || p{1.5cm}<{\centering} | p{1.5cm}<{\centering} || p{2.0cm}<{\centering}}
\toprule
\multirow{2}*{\textbf{Automated GUI testing}} & \multicolumn{2}{c||}{\textbf{Bug detection}} & \multicolumn{2}{c||}{\textbf{Average coverage}} & \textbf{Average bug}  \cr 
 & \textbf{\#Bugs} & \textbf{Recall} & \textbf{Activity} & \textbf{Code} & \textbf{detection time} \\
\midrule
\multicolumn{6}{c}{Random-based method}\\
\midrule
Monkey & 11 & 0.16  & 0.28 & 0.26 &  31.2\\ 
Droidbot & 15 & 0.21  & 0. 31 & 0.30 & 33.9 \\ 
Time-Machine & 19 & 0.27  & 0.34 & 0.33 & 29.7 \\   
\midrule
\multicolumn{6}{c}{Model-based method}\\
\midrule
Fastbot & 20 & 0.29  & 0.34 & 0.32 & 30.3 \\ 
Ape & 18 & 0.26  & 0.33 & 0.31 & 32.5 \\ 
ComboDroid & 21 & 0.30  & 0.37 & 0.35 & 35.1 \\ 
\midrule
\multicolumn{6}{c}{Learning-based method}\\
\midrule
Humanoid & 20 & 0.29  & 0.35 & 0.33 & 33.1 \\  
Q-testing & 18 & 0.26  & 0.32 & 0.31 & 29.6 \\ 
\midrule
\multicolumn{6}{c}{LLM-based method}\\
\midrule
AppAgent & 21 & 0.30  & 0.38 & 0.35 & 28.9 \\ 
GPTDroid & 24 & 0.34  & 0.43 & 0.42 & 33.7 \\ 
DroidAgent & 20 & 0.29  & 0.38 & 0.36 & 35.2 \\ 
AUITestAgent & 23 & 0.33  & 0.41 & 0.38 & 32.4 \\ 
VisionDroid & 26 & 0.37  & 0.49 & 0.47 & 28.1 \\  
Guardian & 27 & 0.39  & 0.51 & 0.49 & 27.7 \\ 
\midrule
\multicolumn{6}{c}{Bug-driven method}\\
\midrule
BugHunter & 34 & 0.49  & 0.48 & 0.45 & 16.6 \\ 
\midrule
\textbf{{\tool}} & \textbf{54} & \textbf{0.77}  & \textbf{0.52} & \textbf{0.50} & \textbf{9.3} \\ 
\bottomrule
\end{tabular}
\vspace{0.05in}
\begin{tablenotes}
\scriptsize
\item \textbf{\textit{Notes:}} ``\#Bugs'' is the number of the detected bugs from automated GUI testing tools.  
\end{tablenotes}
\vspace{-0.15in}
\end{table}

We further present the bug detection performance of {\tool} and baselines in Table~\ref{tab:RQ1}. 
Across 42 apps with 70 confirmed crash bugs, {\tool} achieves a recall of 77\% (54/70), significantly outperforming all 15 baselines. 
It achieves 57\% (0.77 vs. 0.49) recall higher compared with the best baseline (BugHunter) across these apps. In terms of the number of detected bugs, {\tool} can still achieve a higher detection rate. We also compare the similarities and differences of the bugs between BugHunter and {\tool}, and the results show that all bugs detected by BugHunter are also detected by {\tool}. This indicates the effectiveness of {\tool} in detecting crash bugs and helps to ensure app quality.

Table~\ref{tab:RQ1} also shows the average bug detection time of {\tool} and the baselines. We can see that {\tool} detects bugs significantly faster than all baselines. The average detection time of {\tool} is 9.3 minutes, while the best baseline BugHunter requires 16.6 minutes, which means {\tool} reduces 44\% (9.3 vs. 16.6) of the time to trigger a crash. This again proves the efficiency of {\tool}, which is valuable for saving more time for bug detection. 
Table~\ref{tab:RQ1} further reports the average activity coverage and code coverage of {\tool} and the baselines. We can see that {\tool} achieves 0.52 activity coverage and 0.50 code coverage across the 42 apps, which is comparable to BugHunter (0.48 and 0.45) and even slightly higher. Considering that {\tool} is a GUI testing tool and the goal of {\tool} is to detect more crash bugs, it can still achieve and exceed the optimal baseline in terms of testing coverage. This also indicates the effectiveness of {\tool} in covering more activities and code regions. 
We further analyze the cost and latency overhead of {\tool} to contextualize the reported detection time. The average detection time reported in Table~\ref{tab:RQ1} measures the elapsed time until a crash is first triggered during dynamic exploration, and does not include the one-time preprocessing overhead. In practice, the static analysis phase takes 0.9-1.8 minutes per app on average, dominated by AST and CFG construction, while the LLM-based reasoning stage introduces an additional latency of 40-70 seconds. As this preprocessing is performed only once before testing and does not scale with the exploration duration, its overhead is amortized over the entire testing process and does not affect the relative efficiency comparison across tools.

The advantages of {\tool} are due to three aspects. 
First, {\tool} identifies suspicious code segments and uses them to guide testing, which helps to focus exploration on error-prone areas. 
Second, {\tool} excels at reconstructing complex triggering paths that involve repeated navigation and backtracking before reaching the crash, which are often missed by coverage-driven baselines. 
Third, the suspicious path-guided execution ensures that these high-risk paths are explored with priority, significantly improving both effectiveness and efficiency in bug detection.

\subsection{Ablation Study (RQ2)}
\label{sec_results_RQ2}
We first study the effect of removing suspicious-code guidance entirely, by disabling the static analysis module and relying only on dynamic execution over the ATG. In this setting, the system explores the app using runtime execution and execution-aware monitoring, but without any suspicious code signals to guide path generation or prioritization. Under this configuration, the recall drops to 0.49, which is substantially lower than the full {\tool} (0.77). This result indicates that dynamic execution alone, even when combined with execution-aware monitoring, is insufficient to reliably uncover crash-inducing behaviors without code-level guidance.

\begin{table}[t]
\vspace{0.05in}
\renewcommand\arraystretch{0.9} 
\caption{Contribution of different sub-modules (RQ2)}
\vspace{-0.05in}
\label{tab:RQ2}
\centering
\footnotesize
\begin{tabular}{p{3.7cm}<{\centering} | p{0.8cm}<{\centering} | p{0.9cm}<{\centering} || p{2.3cm}<{\centering} | p{1.9cm}<{\centering} || p{1.9cm}<{\centering}}
\toprule
\textbf{Sub-module} & \textbf{\#Bugs} & \textbf{Recall} & \textbf{Activity coverage} & \textbf{Code coverage} & \textbf{Detection time}  \\
\midrule
\textbf{{\tool}} & \textbf{54} & \textbf{0.77}  & \textbf{0.52} & \textbf{0.50} & \textbf{9.3} \\ 
\midrule
\textit{w/o Interaction Signals Collection} & 44 & 0.63 & 0.47 & 0.46 & 11.4 \\
\textit{w/o Rough Path Generation} & 47 & 0.67 & 0.47 & 0.45 & 12.2\\
\textit{w/o Executable Path Completion} & 41 & 0.59 & 0.43 & 0.42 & 14.7 \\
\textit{w/o Global Path Planning} & 42 & 0.60 & 0.47 & 0.46 & 11.1\\
\textit{w/o Execution aware Monitoring} & 43 & 0.61 & 0.45 & 0.44 & 11.8\\
\bottomrule
\end{tabular}
\vspace{-0.15in}
\end{table}

Table~\ref{tab:RQ2} further shows the performance of {\tool} and its five variants, where each variant removes one key sub-module. 
For \textit{w/o Interaction Signals Collection}, we remove the backtracking of suspicious code segments to their triggering callbacks and upstream operations, meaning suspicious code is only linked to widgets without considering interaction signals.  
For \textit{w/o Rough Path Generation}, we disable the matching process and instead rely only on direct ID matching, which ignores cases where IDs are obfuscated or missing.  
For \textit{w/o Executable Path Completion}, we remove the LLM-based completion, leaving only the raw ATG and static mappings without path refinement or disambiguation.  
For \textit{w/o Global Path Planning}, we disable the LLM-based global scheduling and simply traverse the ATG in a breadth-first manner.  
For \textit{w/o Execution-aware Monitoring}, we disable runtime monitoring of execution states and crashes, so the system only records completed paths without detecting incomplete or abnormal terminations.

From Table~\ref{tab:RQ2}, we can see that {\tool}'s bug detection recall, activity coverage, and code coverage are consistently higher than all other variants, demonstrating the necessity of each designed sub-module.  
Compared with {\tool}, removing the LLM-based path completion (\textit{w/o Executable Path Completion}) leads to the largest performance decline, i.e., 23\% drop in recall (0.59 vs. 0.77) and 8-9\% decline in coverage. This highlights the importance of context-aware reasoning to complete missing paths and consolidate noisy mappings.  
Removing rough path generation (\textit{w/o Rough Path Generation}) also causes a noticeable decrease in recall, i.e., 13\% drop (0.67 vs. 0.77), since many widgets cannot be accurately resolved by ID alone.  
Code-level interaction signals (\textit{w/o Interaction Signals Collection}) also contribute significantly, with an 18\% recall reduction (0.63 vs. 0.77), because without tracing upstream callbacks, suspicious code often cannot be tied to the correct triggering operations.  
Removing global path planning (\textit{w/o Global Path Planning}) decreases recall by 22\% (0.60 vs. 0.77), showing that the LLM-guided scheduling effectively reduces redundant exploration and prioritizes high-risk paths.  
Finally, removing execution-aware monitoring (\textit{w/o Execution-aware Monitoring}) causes a 21\% drop (0.61 vs. 0.77), since the tool cannot distinguish between successful and incomplete executions, leading to missed detections. These results confirm that each sub-module contributes to the overall effectiveness of {\tool}. The LLM-based executable path completion is particularly crucial, as it bridges the gap between partial static analysis results and runtime execution. Meanwhile, the other modules together ensure accurate signal extraction, path generation, efficient scheduling, and reliable detection of crash-inducing behaviors.

\begin{table}[t]
\renewcommand\arraystretch{0.9} 
\caption{Impact of different large language models on bug detection performance (RQ2)}
\vspace{-0.05in}
\label{tab:RQ2-2}
\centering
\footnotesize
\begin{tabular}{p{3.7cm}<{\centering} | p{0.8cm}<{\centering} | p{0.9cm}<{\centering} || p{2.3cm}<{\centering} | p{1.9cm}<{\centering} || p{1.9cm}<{\centering}}
\toprule
\textbf{Different LLM} & \textbf{\#Bugs} & \textbf{Recall} & \textbf{Activity coverage} & \textbf{Code coverage} & \textbf{Detection time}  \\
\midrule
\textbf{{\tool} (GPT-4o)} & \textbf{54} & \textbf{0.77}  & \textbf{0.52} & \textbf{0.50} & \textbf{9.3} \\ 
\midrule
\textit{GPT-4o mini} & 48 & 0.69 & 0.47 & 0.46 & 12.4 \\
\textit{Gemini-1.5-Pro} & 51 & 0.73 & 0.50 & 0.48 & 10.5\\
\textit{LLaMA-4} & 50 & 0.71 & 0.49 & 0.48 & 10.9\\
\textit{Qwen2.5-72B} & 47 & 0.67 & 0.47 & 0.46 & 12.7 \\
\bottomrule
\end{tabular}
\vspace{-0.15in}
\end{table}

We further investigate the impact of different large language models on the performance of {\tool}. As shown in Table~\ref{tab:RQ2-2}, {\tool} achieves the best performance when using GPT-4o, while still maintaining competitive recall when paired with other LLMs.  Gemini-1.5-Pro and LLaMA-4 achieve recalls of 0.73 and 0.71, with comparable coverage, respectively, demonstrating that the proposed framework is compatible with different LLMs. We observe that weaker models are more likely to generate invalid or partially inconsistent interaction sequences. In contrast, GPT-4o shows stronger reasoning capability over mixed structural inputs and more stable adherence to the required output schema.

\subsection{Usefulness Evaluation (RQ3)}
\label{sec_results_RQ3}
Based on the results of RQ1 in Section~\ref{sec_results_RQ1}, we select the baselines with the best bug detection performance in each category (Time-Machine, ComboDroid, Humanoid, Guardian, BugHunter) to run on a set of real-world popular apps. {\tool} and the baseline methods are configured with the same running time, with each app running for 60 minutes.

For this experiment, we test 116 apps downloaded from Google Play, and {\tool} successfully detects 21 previously unknown crash bugs across these apps. Among these, 15 bugs have already been fixed by developers, and 6 bugs have been confirmed and marked for future fixes. In contrast, the best-performing baseline (BugHunter) detects only 6 of these bugs, and all are a strict subset of those identified by {\tool}. None of the other baselines find any additional bugs beyond those already covered by {\tool}. We further inspect the triggering paths of these newly discovered crashes and find that they typically require multi-step interactions, with an average length of 8.7 GUI actions.
This again highlights the effectiveness of {\tool} in exposing crash-inducing behaviors that remain hidden from existing automated GUI testing. 

\begin{table}[htb]
\renewcommand\arraystretch{0.9} 
\caption{Information on the confirmed or fixed bugs. (RQ3)}
\vspace{-0.05in}
\label{tab:RQ3}
\centering
\scriptsize
\begin{tabular}{p{0.2cm}<{\centering} | p{1.6cm}<{\centering} | p{0.6cm}<{\centering} | p{1.0cm}<{\centering} | p{0.8cm}<{\centering} | p{1.0cm}<{\centering} | p{0.5cm}<{\centering} | p{0.6cm}<{\centering} | p{0.6cm}<{\centering} | p{0.6cm}<{\centering} | p{0.5cm}<{\centering} | p{1.4cm}<{\centering}}
\toprule
\textbf{ID} & \textbf{App name}  & \textbf{Down} & \textbf{Category} & \textbf{Version} & \textbf{Status} & \textbf{Time} & \textbf{Comb} & \textbf{Huma} & \textbf{Guard} & \textbf{BugH} & \textbf{Bug Type}\\
 \midrule
1 &  Journal &  5M+ &  Product &  131 &  Fixed &   &   &   &   &  & NullPointer\\
2 &  Height Increase &  5M+ &  Fitness &  1.1.8 &  Fixed &   &   &   &   &  & RuntimeError\\
3 &  Offline Music &  5M+ &  Music &  1.34.1 &  Fixed &   &   &   & $\ast$ &  $\ast$& NullPointer\\
4 &  DailyLife &  5M+ &  Lifestyle &  4.3.1 &  Fixed &   &   &   &   &  & ExternalInte\\
5 &  Easy Notes &  1M+ &  Product &  1.3.23 &  Fixed &   &  $\ast$ &   &   &  $\ast$& RuntimeError\\
6 &  ExpenseTracker &  1M+ &  Finance &  1.01 &  Fixed &   &   &   &   &  & Concurrency\\
7 &  My Cash &  1M+ &  Enter &  38 &  Confirmed &   &   &   &  $\ast$ &  & NullPointer\\
8 &  BetterMe &  1M+ &  Health &  9.31 &  Fixed &   &  $\ast$ &   &   &  $\ast$& DBOperation\\
9 &  Daily Yoga &  1M+ &  Sport &  1.2.1 &  Fixed &   &   &   &   &  & NullPointer\\
10 &  Daily Routine &  1M+ &  Tool &  1.1.10 &  Confirmed &   &   &   &   &  & RuntimeError\\
11 &  FreeYourMusic &  1M+ &  Music &  9.17.0 &  Fixed &   &   &  $\ast$ &   &  & NullPointer\\
12 &  WeatherForecast &  1M+ &  Weather &  2.1 &  Confirmed &   &   &   &   &  $\ast$& RuntimeError\\
13 &  Add Text &  1M+ &  Photo &  2.7 &  Fixed &  $\ast$ &   &   &  $\ast$ &  & NullPointer\\
14 &  SideChef &  1M+ &  Food &  5.31.1 &  Fixed &   &   &   &   &  & ExternalInte\\
15 &  RISE &  1M+ &  Lifestyle &  1.78.33 &  Fixed &   &   &   &   &  & ExternalInte\\
16 &  Timetable &  1M+ &  Education &  2.3.1 &  Confirmed &   &   &   &   &  & Concurrency\\
17 &  Daily Planner &  1M+ &  Product &  63 &  Fixed &  $\ast$ &  $\ast$ &  $\ast$ &  $\ast$ &  $\ast$& RuntimeError\\
18 &  Headout &  1M+ &  Travel &  8.19.0 &  Confirmed &   &   &   &   &  & ExternalInte\\
19 &  MyMoney &  1M+ &  Finance &  5.7 &  Fixed &   &   &   &   &  & DBOperation\\
20 &  Quabble &  500K+ &  Health &  2.6.8 &  Confirmed &   &   &   &   &  $\ast$& Concurrency\\
21 &  EdrawMind &  500K+ &  Tool &  7.9.1 &  Fixed &   &   &   &   &  & DBOperation\\
\bottomrule

\end{tabular}
\vspace{-0.1in}
\end{table}

Table~\ref{tab:RQ3} lists the bugs that are confirmed of fixed by developers, the screenshot of the email is available on our website\textsuperscript{\ref{github}}. The responses from developers further confirm the usefulness of {\tool}. One developer replied: ``\textit{Thanks, we were not aware of this bug. It was due to improper handling of null objects in the background service.}'' 
These results demonstrate that {\tool} not only improves bug detection performance under open source apps but also provides practical value in real-world popular apps.

\begin{table}[htb]
\vspace{-0.05in}
\renewcommand\arraystretch{1} 
\caption{Comparison of TraceDroid with mainstream static analysis tools in bug detection}
\vspace{-0.05in}
\label{tab:RQ-diss-1}
\centering
\footnotesize
\begin{tabular}{p{1.3cm}<{\centering} || p{1.7cm}<{\centering} | p{1.4cm}<{\centering} | p{1.0cm}<{\centering} | p{1.0cm}<{\centering} || p{1.8cm}<{\centering}}
\toprule
\textbf{Metric} & Android Lint & SpotBugs & PMD & Infer & \textbf{{\tool}} \\
 \midrule
\textbf{\#Bugs} & 8 & 10 & 13 & 10 & \textbf{54} \\
\textbf{Recall} & 0.11  & 0.14  & 0.19  & 0.14 & \textbf{0.77} \\
\bottomrule
\end{tabular}
\vspace{-0.1in}
\end{table}

\section{Discussion}
\label{sec_discussion}

\subsection{Compatibility with Static Analysis Tools}
The first sub-module of {\tool} (Heuristic-based Suspicious Code Detection) shares some similarities with traditional static analysis tools, Android Lint~\cite{lint}, SpotBugs~\cite{SpotBugs}, PMD~\cite{PMD}, and Infer~\cite{Infer}, which are also designed to scan code and highlight abnormal patterns. To evaluate whether our subsequent modules can enhance the findings of these tools, we replace the first module of {\tool} with each of them and then run the rest of our pipeline unchanged. 
As shown in Table \ref{tab:RQ-diss-1}, traditional static analysis tools alone detect very few crash bugs, with recalls of 0.11, 0.14, 0.19, and 0.14, respectively. In contrast, {\tool} achieves a recall of 0.77, which is 58\% higher than the best-performing static analysis baseline (PMD). The gap stems from the fact that these tools primarily flag isolated code issues without considering how they can be triggered at runtime. By feeding their outputs into our second and third modules, however, {\tool} is able to map suspicious segments to GUI components, reconstruct feasible triggering paths, and validate them through dynamic execution. This demonstrates that our framework not only outperforms existing static analysis tools when used alone but can also serve as a general enhancer to amplify their findings in the context of automated GUI testing.

\subsection{Enhancing LLM-Based GUI Testing with {\tool}}
To integrate {\tool} with existing LLM-based testers (the ``+'' variants), we inject the suspicious paths generated by {\tool} into their prompts. Specifically, instead of replacing their decision-making process, we prepend an additional instruction to the prompt, stating that following the given suspicious paths may help uncover crash bugs, and explicitly list the paths. 
With this integration, we observe clear improvements across all baselines. The most improvement is observed in Guardian, whose recall rises from 0.39 to 0.51, representing a 31\% relative gain. This demonstrates that {\tool} complements LLM-based methods by injecting targeted suspicious paths into their prompts, thereby reducing exploration randomness and steering them toward higher-value interactions. {\tool} not only outperforms standalone baselines, but it can also act as a general enhancement layer to strengthen existing LLM-based GUI testing frameworks.

\begin{table}[htb]
\renewcommand\arraystretch{1} 
\caption{Performance of integrating TraceDroid into existing LLM-based GUI testing methods}
\vspace{-0.05in}
\label{tab:RQ-diss-2}
\centering
\footnotesize
\begin{tabular}{p{0.7cm}<{\centering} || p{1.3cm}<{\centering} | p{0.6cm}<{\centering} | p{0.6cm}<{\centering} |  p{0.6cm}<{\centering} | p{0.6cm}<{\centering} | p{0.6cm}<{\centering} | p{0.6cm}<{\centering} || p{0.6cm}<{\centering} | p{0.5cm}<{\centering} | p{0.5cm}<{\centering} | p{0.5cm}<{\centering} | p{0.5cm}<{\centering} | p{0.5cm}<{\centering}}
\toprule
\multirow{2}*{\textbf{Metric}}  & \multirow{2}*{\textbf{{\tool}}} & \multicolumn{6}{c||}{\textbf{Tools with {\tool}}} & \multicolumn{6}{c}{\textbf{Tools without {\tool}}} \cr 
 &  & \textbf{ApA+} & \textbf{GPD+} & \textbf{DRA+} & \textbf{AUI+} & \textbf{VID+} & \textbf{GUA+}  & \textbf{ApA} & \textbf{GPD} & \textbf{DRA} & \textbf{AUI} & \textbf{VID} & \textbf{GUA}\\
 \midrule
\textbf{\#Bugs} & \textbf{54} & 24 & 27 & 26 & 29 & 34 & 36 & 21 & 24 & 20 & 23 & 26 & 27 \\
\textbf{Recall} & \textbf{0.77}  & 0.34  & 0.39  & 0.37  & 0.41  & 0.49  & 0.51 & 0.30 & 0.34 & 0.29 & 0.33 & 0.37 & 0.39 \\
\bottomrule
\end{tabular}
\vspace{0.05in}
\begin{tablenotes}
\scriptsize
\item \textbf{\textit{Notes:}} ``ApA'' is the AppAgent, ``GPD'' is the GPTDroid, ``DRA'' is the DroidAgent, ``AUI'' is the AUITestAgent, ``VID'' is the VisionDroid, ``GUA'' is the Guardian.  
\end{tablenotes}
\vspace{-0.15in}
\end{table}

\subsection{Advantages over Test Case Migration and Bug Replay Techniques}
Test case migration (e.g., AppTestMigrator~\cite{behrang2019test}, Craftdroid~\cite{lin2019test}, TEMdroid~\cite{zhang2024learning}) and bug replay techniques (e.g., RecDroid+\cite{zhao2022recdroid+}, ScopeDroid~\cite{huang2023context}, AdbGPT~\cite{feng2024prompting}) are designed to reuse existing execution traces or adapt historical test cases to new app versions. For a fair comparison, we adapt {\tool}'s outputs to the input formats expected by these techniques. Specifically, the suspicious paths generated by our context-aware path generation module are rewritten into migration-style test cases and replay-style step sequences, allowing existing tools to consume them as if they were historical traces. As shown in Table \ref{tab:RQ-diss-3}, {\tool} detects 54 crash bugs with a recall of 0.77, substantially outperforming all migration and replay methods, whose recalls remain below 0.31. Even Craftdroid and TEMdroid, detect only 22 bugs (0.31 recall), representing less than half of {\tool}'s detection capability. This demonstrates that directly reasoning from suspicious code-level signals to executable paths provides a more proactive and effective means of discovering crashes than passively adapting past traces. 

\begin{table}[htb]
\vspace{-0.05in}
\renewcommand\arraystretch{1} 
\caption{Performance of different test case migration and bug replay methods}
\vspace{-0.05in}
\label{tab:RQ-diss-3}
\centering
\footnotesize
\begin{tabular}{p{1.3cm}<{\centering} || p{1.4cm}<{\centering} | p{1.4cm}<{\centering} | p{1.0cm}<{\centering}  || p{1.8cm}<{\centering} | p{1.3cm}<{\centering} | p{1.3cm}<{\centering}  || p{1.3cm}<{\centering}}
\toprule
\textbf{Metric} & RecDroid+ & ScopeDroid & AdbGPT & AppTestMigrator & Craftdroid & TEMdroid & \textbf{{\tool}} \\
 \midrule
\textbf{\#Bugs} & 18 & 19 & 17 & 20 & 22 & 22 & \textbf{54} \\
\textbf{Recall} & 0.26  & 0.27  & 0.24  & 0.29  & 0.31  & 0.31  & \textbf{0.77} \\
\bottomrule
\end{tabular}
\vspace{-0.1in}
\end{table}

\subsection{Threats of Validity}
\label{subsec_Validity}
The first threat concerns the representativeness of the studied apps and bugs. In RQ1 and RQ2, we used 42 apps with 70 confirmed crash bugs, and in RQ3 we selected additional popular Google Play apps. Although these datasets cover diverse categories, some domains or crash patterns may still be underrepresented.
The second threat is the potential influence of LLM pretraining. Since LLMs are trained on large corpora, some apps or code snippets similar to ours may have appeared in training. To mitigate this, we excluded apps from our pilot study and chose recently updated Google Play apps.
The third threat relates to LLM hallucinations when generating paths or actions. LLMs may produce non-existent widgets or infeasible transitions. {\tool} employs lightweight validators after each generation step: ATG consistency, widget existence, and action schema, discarding or repairing invalid outputs. 
These safeguards substantially reduce the impact of hallucinations, but cannot completely eliminate the risk of subtle semantic errors.
The fourth threat concerns the completeness of the statically constructed Activity Transition Graph (ATG) and its interaction with LLM-based path generation.
Some dynamic transitions, such as those introduced via reflection-based routing or data-dependent implicit intents, may not be fully captured by static analysis.
In our evaluation, 92\% of inter-activity transitions observed during dynamic execution are already present in the ATG, while the remaining 8\% fall into these hard-to-analyze cases.
For LLM-generated paths, 14\% require at least one deterministic repair (e.g., replacing an unreachable hop with a shortest valid path in the ATG), and 6\% are discarded due to invalid widget references.
These cases are effectively handled by validation and repair mechanisms before execution, but extreme cases of highly dynamic navigation may still pose challenges.

\section{Related Work}
\label{sec_related}
\textbf{Automated GUI Testing.}
To ensure the quality of mobile apps, many researchers study the automatic generation of large-scale test scripts to test apps~\cite{xie2007designing}.
Since Android apps are event-based~\cite{anand2012automated,wu2019analyses,jabbarvand2019search,matinnejad2017automated}, the most common automated testing approaches are model-based automated GUI testing approaches~\cite{mirzaei2016reducing,yang2018static,yang2013grey,zeng2016automated,mao2016sapienz,su2017guided,dong2020time,gu2019practical,wang2020combodroid}, which design corresponding models through the analysis of the apps.
Due to the lack of consideration for the semantic information of the app's GUI pages, the coverage of model-based approaches is still low.
Researchers further proposed human-like testing strategies and designed LLM-based automated GUI testing approaches. GPTDroid~\cite{liu2023chatting} used GPT-3.5 to generate the testing script. AppAgent~\cite{yang2023appagent} enabled the agent to operate apps through a simplified action space.
DroidAgent~\cite{yoon2023autonomous} generated high-level, realistic tasks for GUI testing based on app-specific functionalities.
AUITestAgent~\cite{hu2024auitestagent} is used for step-oriented testing. AXNav~\cite{taeb2024axnav} generated annotated videos to visually and interactively review accessibility test results.
UXAgent~\cite{lu2025uxagent} achieved LLM-driven automated usability testing by simulating thousands of user interactions. TestAgent~\cite{li2024test} generated and executed test cases through multimodal perception.
Although these methods improve coverage and interaction diversity, they remain limited in efficiently detecting crash-inducing behaviors.

\textbf{Static Program Analysis.}
Static analysis~\cite{yanmultiple,samhi2024call,qiu2018analyzing,mahmud2022acid,dilhara2018automated,wu2021program,samhi2022difuzer,li2016droidra,yang2015static,li2015iccta,pauck2021jicer,onwuzurike2019mamadroid,li2022cross,samhi2021raicc,karakaya2021sootfx} is another widely adopted technique to improve software reliability. Researchers have developed tools targeting different categories of bugs. For null pointer dereference~\cite{tomassi2021real}, Infer~\cite{Infer} and NullAway~\cite{NullAway} analyzed type annotations and data-flow conditions to detect potential unsafe accesses. For concurrency errors~\cite{da2010effective,hong2014detecting,barros2023experimental}, RacerD~\cite{2018RacerD} identified race conditions and missing synchronization in shared-memory programs. For resource management~\cite{brocanelli2018improving,cui2023safedrop,jiang2017detecting,wang2025boosting} and API misuse~\cite{amann2018systematic,li2021large,amann2018evaluation,mahmud2022acid}, FindBugs~\cite{Findbugs} and SpotBugs~\cite{SpotBugs} provide rule-based checks to detect memory leaks, lifecycle misuses, and incorrect Android API calls.

\section{Conclusion}
\label{sec_conclusion}
Detecting crash bugs in mobile apps remains a challenging problem, as many crashes require precise GUI interaction sequences that are difficult to reach with existing automated testing methods. In this paper, we propose {\tool}, a novel automated GUI testing approach that leverages suspicious code-level signals to guide exploration toward crash-prone interactions. {\tool} first detects suspicious code segments and associates them with GUI widgets to collect code-level interaction signals. It then constructs the ATG, generates rough paths, and employs LLM-based executable path completion to anchor suspicious interactions to feasible execution routes. Finally, {\tool} executes these suspicious paths through global path planning, local path generation, and execution-aware monitoring, enabling efficient and reliable crash detection.
The experimental results demonstrate the effectiveness of {\tool}. These results confirm that {\tool} can effectively transform suspicious static traces into actionable execution paths, improving both the efficiency and practicality of automated GUI testing.

In the future, we plan to extend {\tool} to cover a wider range of abnormal behaviors beyond crashes, and to investigate tighter integration with emerging LLM-based GUI testing methods for further enhancing its adaptability and bug detection capability.

\section{Data Availability}
\label{sec_Data Availability}
We release the source code, dataset, and experimental results on our website \url{https://github.com/TheMystery123/TraceDroid}.

\begin{acks}
This work was supported by the National Natural Science Foundation of China Grant No. 62402483, 62232016, 62072442, 62402484, Youth Innovation Promotion Association CAS, Basic Research Program of ISCAS Grant No. ISCAS-JCZD-202304, Major Program of ISCAS Grant No. ISCAS-ZD- 202302 and ISCAS-ZD-202401, Innovation Team 2024 ISCAS (No. 2024-66).
\end{acks}

\bibliographystyle{ACM-Reference-Format}
\bibliography{reference}

\end{document}